\documentclass[
twocolumn,
groupedaddress,
notitlepage
]
{revtex4-1}
\RequirePackage{amsmath,amssymb,bm,wasysym}
\RequirePackage{graphicx}
\RequirePackage{hyperref}
\usepackage{siunitx}
\usepackage{tikz}
\usepackage{subcaption}
\usepackage{blkarray}

\hypersetup{
breaklinks={true},
citecolor={blue},
colorlinks={true},
linkcolor={blue},
pdfcreator={\LaTeX and\flqq hyperref\frqq},
pdffitwindow={true},
pdfmenubar={true},
pdfpagelayout={OneColumn},
pdfstartview={FitH},
pdftoolbar={true},
plainpages={false},
pdfpagemode={UseThumbs},
}
\usepackage{braket}
\DeclareMathAlphabet{\mathpzc}{OT1}{pzc}{m}{it}

\newcommand{\minus}{\scalebox{0.45}[1.0]{$-$}}

\begin{document}
\pagestyle{plain}
\title{Mean Field Approximation for solving QUBO problems}
\author{M\'at\'e Tibor Veszeli}
\affiliation{Institute of Physics, E\"{o}tv\"{o}s University, 1518 Budapest, Hungary}
\author{G\'abor Vattay}
\affiliation{Institute of Physics, E\"{o}tv\"{o}s University, 1518 Budapest, Hungary}
\pacs{
}
\begin{abstract}

The Quadratic Unconstrained Binary Optimization (QUBO) problems are NP hard; thus, so far, there are no algorithms to solve them efficiently. There are exact methods like the Branch-and-Bound algorithm for smaller problems, and for larger ones, many good approximations like stochastic simulated annealing for discrete variables or the mean field annealing for continuous variables. This paper will show that the statistical physics approach and the quantum mechanical approach in the mean field annealing give the same result. We examined the Ising problem, which is an alternative formulation of the QUBO problem. Our methods consist of a set of simple gradient-based minimizations with continuous variables, thus easy to simulate. 
We benchmarked our methods with solving the Maximum Cut problem with the G-sets. In many graphs, we could achieve the best-known Cut Value. 

\end{abstract}
\maketitle

\section{Introduction}
\label{sec:int}
Spin models are versatile because they are simple yet able to demonstrate fundamental phenomenons, like phase transition \cite{Erns-1925, PhysRev.65.117, Baxter-stat_mecha}. Many complex physical models can be reduced to a simple Ising or Heisenberg model, like electron and nuclear spins \cite{solyom2007fundamentals}, and even social situations \cite{mezard1987spin}.
It is also important in modern applied physics since many
real-life problems can be traced back to find the global minimum of a high-dimensional, nonlinear function. Most of these tasks are NP-hard \cite{barahona1982computational}, thus there exist no effective method to solve them, but there are many good numerical approximations, like the stochastic simulated annealing \cite{kirkpatrick1983optimization, Cerny1985, isakov2015optimised}, mean field annealing \cite{bilbro1988optimization}, tabu search \cite{kochenberger2013solving, glover2010diversification}, semidefinite programming \cite{rendl2010solving, goemans1995improved, poljak1995solving}, and special devices like coherent Ising machine \cite{inagaki2016coherent, haribara2016coherent}, adiabatic quantum computer \cite{steffen2003experimental,johnson2011quantum} to treat them, like the D-Wave system \cite{harris2018phase, farhi2000quantum, roland2002quantum, hamerly2019experimental}
There are also exact methods like the Branch-and-Bound \cite{rendl2007branch}, or Branch-and-Cut \cite{padberg1991branch} algorithms, but their drawback is they cannot handle too many nodes. In a dense problem, approximately up to 100 nodes. 

In the current paper, we introduce a mean field approximation based algorithm for solving the QUBO problem. The structure of this paper is the following.
In section \ref{sec:ising_and_qubo} we summarize the definition of the Ising model and the QUBO problems shortly and show their connection.
In section \ref{sec:var_princ} and \ref{sec:quantum_mechanical_approach} we present two equivalent methods to give a good result to the Ising and QUBO problem. The first builds on the variational principle of statistical physics \cite{peierls1938minimum, opper2001advanced} with annealing, the second on the variational principle of quantum mechanics \cite{griffiths2018introduction} with the adiabatic theorem \cite{Born1928, Kato1950}. Finally in section \ref{sec:benchmark} we benchmark our program with the G-sets \cite{helmberg2000spectral}.

\section{Ising model and QUBO problems}
\label{sec:ising_and_qubo}
The Ising model consist of interacting spins: $\underline{S} = (S_1, S_2 \dots S_N)$, with components $S_i \in \{ \pm 1 \}$. The model is defined by its energy:
\begin{equation}
\label{eq:ising_def}
E_{\underline{S}} = - \frac{ 1 }{ 2 } \sum_{i j} J_{i j} S_i S_j
- \sum_i h_i S_j,
\end{equation}
where $J_{i j}$ is the interaction between spin $i$ and $j$, and $h_i$ is the external magnetic field. We assume, that $J_{i j} = J_{j i}$ and $J_{i i} = 0$. A relevant question is what is the critical temperature, what are the temperature dependencies of the expected values or correlation, and if the system is frustrated than even the ground state is nontrivial.

A QUBO problem is defined by
\begin{equation}
\underset{ \underline{x} }{ \mathrm{argmin} } \{ q(\underline{x}) \},
\end{equation}
where
\begin{equation}
\label{eq:QUBO-def}
q(\underline{x}) = \sum_{i j} Q_{ij} x_i x_j \qquad x_i \in \{0,1\},
\end{equation}
and $\underline{\underline{Q}}$ is a symmetric matrix. Substituting $x_i = (1+S_i)/2$ in equation \ref{eq:QUBO-def} yields
\begin{equation}
q = \frac{1}{4} \sum_{ \underset{(i \neq j)}{i j} } Q_{i j} S_i S_j
+ \frac{1}{2} \sum_i \left( \sum_j Q_{i j} \right) S_i
+ \mathrm{const}
\end{equation}
which means if $J_{i j} = - \frac{ 1 }{ 2 } Q_{i j}$ for $i \neq j$, $J_{i i}=0$ and $h_i = -\frac{1}{2} \sum_j Q_{i j}$ then the QUBO problem is equivalent to finding the ground state of the Ising model. As it was summarized by Lucas, many NP hard problem can be formulated with the Ising model \cite{lucas2014ising}

One typical QUBO problem is the Maximum Cut problem \cite{karp1972reducibility}.
The task is to partition an undirected graph ($\mathcal{G} = (\mathcal{V}, \mathcal{E})$) into two subsets ($\mathcal{S}, \mathcal{V} \backslash \mathcal{S}$) such as that the number of edges between these subsets is as large as possible. If the graph is defined via its adjacency matrix ($W_{ij}$), than the corresponding cut value (CV) is
\begin{equation}
\label{eq:cut_value}
\text{CV} = 
\sum_{ \{i, j\} \in \mathcal{E},
i \in \mathcal{S}, j \in \mathcal{V} \backslash \mathcal{S} }
W_{i j}
= \sum_{ \underset{(i < j)}{i,j} } W_{i j} \dfrac{ 1 - S_i S_j }{ 2 }.
\end{equation}
In equation \ref{eq:cut_value} $S_i$ is the spin variable and $S_i = 1\ (\uparrow)$ means the $i$th spin is in the subset $\mathcal{S}$, and $S_i = -1\ (\downarrow)$ means it's in $\mathcal{V} \backslash \mathcal{S}$. Maximizing the cut value is equivalent to minimize the Ising energy, where $J_{i j} = - W_{i j}$.

\section{ Variational approach in statistical physics }
\label{sec:var_princ}

The variational principle of statistical physics is a powerful tool to examine interactive systems at finite temperatures. The most straightforward version of this principle is the mean field approximation, albeit not perfect, but simple, suggestive, and works fine if the number of links per node (spin) is large enough. E.g., the mean field approximation predicts phase transition even at the one-dimensional Ising model, since we know that this is false, but as we increase the dimension, the exact critical temperature and the mean field critical temperature approach, and in the uniform fully connected Ising model the two temperatures are the same.
In a real-life, nonphysical problem, the number of nodes usually not too large (hundreds or thousands but not $10^{23}$ ), but there is no symmetry. As the number of links per node increases, the mean field approximation improves.

One of the central quantities we are interested in statistical physics at finite temperature is free energy.
\begin{equation}
\label{eq:exact_free_energy}
F^\text{(exact)}(T) = -T \ln\left( \sum_n \mathrm{e}^{ - \frac{E_n}{T} } \right)
\end{equation}
Here $T$ is the temperature, the Boltzmann factor is 1, the sum goes over all the states of the system, and $\{E_n\}$s are the energies.
The probability of finding the system at state $n$ is $P^\text{(e)}_n \propto \mathrm{e}^{ - E_n / T }$ therefore at zero temperature the free energy is the ground state energy. The problem with equation (\ref{eq:exact_free_energy}) is that even if we know all the $E_n$ energies, apart from the simplest cases we can't evaluate the summation.
The variational principle states that the exact free energy is always smaller or equal to the variational free energy: $F^\text{(e)}(T) \leq F(T)$.
The variational free energy is
\begin{equation}
\label{eq:variatonal_free_energy}
F(T) = \langle E_n \rangle - T S_\text{inf} = \sum_n P_n E_n
+ T \sum_n P_n \ln(P_n),
\end{equation}
where $\langle E_n \rangle$ is the energy average, $S_\text{inf}$ is the information entropy and $P_n$ can be any probability distribution, but the better our guess, the lower $F(T)$ will be. We have to take account, that in practice the distribution can't be too difficult because we have to calculate analytically the variational free energy, otherwise it is futile. The typical strategy is to consider a class of probability distribution with parameters $\underline{a}$: $P_n(\underline{a})$, calculate $F(\underline{a}, T)$ and finally minimize in $\underline{a}$. This solution is temperature dependent and we will refer to as $\underline{a}(T)$. This $\underline{a}$ parameter might have physical meaning e.g. averages. The variational free energy is now $F(T) = F(\underline{a}(T),T)$.

\subsection{Mean field annealing}
Stochastic simulated annealing, e.g., the idea of imitating the annealing of materials, is well known in computer science \cite{kirkpatrick1983optimization}. However, in that case, the jumps are between discrete states, and now we have continuous parameters. 

In general the $F(\underline{a}, T)$ function has more than one minimum. The deeper minimum we can find, the closer we are to the exact free energy. Nevertheless, it is technically impossible to find all the minima in a complex system. Using a random point in the phase space of $\underline{a}$ and then with some gradient method finding a local minimum gives us typically a bad minimum. A better strategy is to determine $\underline{a}(T)$ at high temperature, where it is easy since only the entropic term is dominant, and then gradually decrease $T$ by $\Delta T$ and find the new $\underline{a}(T - \Delta T)$ which is close to $\underline{a}(T)$. Repeating this procedure will lead to a low-temperature solution where the energy term is dominant. The whole $T \mapsto \underline{a}(T)$ function is the trajectory.

Nothing guarantees that the final solution will be the one with the lowest free energy. Initially, there is only one solution, but during the cooling, more and more can emerge. These minima also move continuously, and at some point, some of them can be smaller than $\underline{a}(T)$.

In the mean field approximation, we assume that a variational distribution factorizes. In the case of spin systems
\begin{equation}
P^\text{MF}({\underline{S}}; \underline{m}) = \prod_{i=1}^N P_i(S_i; m_i) =
\prod_i \frac{ 1 + m_i S_i }{ 2 }
\end{equation}
This distribution is normed ($ \sum_{\underline{S}} P^\text{MF}(\underline{S}) = 1 $) and the expected value is simply $\langle S_i \rangle = m_i$. The variational free energy is
\begin{equation}
\begin{aligned}
\label{eq:mean_field_free_energy}
F^\text{MF}(\underline{m}, T) = - \frac{ 1 }{ 2 } \sum_{i j} J_{i j} m_i m_j
- \sum_i h_i m_j\\
+ T
\sum_i \Big[ \dfrac{1+m_i}{2} \ln\left( \frac{1+m_i}{2} \right)\\
+\dfrac{1-m_i}{2} \ln\left( \frac{1-m_i}{2} \right)
\Big].
\end{aligned}
\end{equation}
This function has to be minimal, so its derivative is zero
\begin{equation}
\label{eq:mean_field_gradient}
\frac{ \partial F^\text{MF} }{ \partial m_i } =
-\sum_j J_{i j} m_j - h_i 
+T \frac{1}{2} \ln\left( \frac{ 1+m_i }{ 1-m_i } \right) = 
0
\end{equation}
and the second derivative, the Hesse matrix is positive definite
\begin{equation}
\label{eq:mean_field_hesse}
\frac{ \partial^2 F^\text{MF} }{ \partial m_i \partial m_j } = - J_{i j} + \frac{ T \delta_{i j} }{ 1-m_i^2 } \succ 0.
\end{equation}
Equation \ref{eq:mean_field_gradient} is the equation of state, which is an implicit equation. We can also formulate a self-consistent equation:
\begin{equation}
\label{eq:mean_field_self_consistent}
m_i = \tanh\left( \frac{1}{T}\left( h_i + \sum_j J_{i j} m_j \right) \right)
\end{equation}
At high temperatures, the solution of this equation is $m_i = h_i/T$ $\forall i$, which is close to 0, but at lower temperatures, there can be more than one solution.
If $\underline{m}(T)$ is known one procedure to determine $\underline{m}(T-\Delta T)$ is to use the self-consistent equation iteratively with the initial guess $\underline{m}(T)$. This method requires the fewest function evaluations, but in many cases, under a certain temperature, $T^*$ the iteration will not converge but oscillate between two values. This temperature is usually not the critical temperature.
The critical temperature $T_\text{c}$ is only well defined if there is no external magnetic field. In this $\underline{h} = 0$ case the $\underline{m} = 0$ solution is a minimum as long as all the eigenvalues of the $\partial_i \partial_j F^\text{MF}( \underline{m}=0 )$ matrix from equation \ref{eq:mean_field_hesse} are positive. That concludes $T_\text{c} = \max_i( \lambda_i(\underline{\underline{J}}) )$. On the other hand equation \ref{eq:mean_field_self_consistent} has the form $\underline{m} = \underline{f}(\underline{m})$, where the $\underline{m} = 0$ solution is an attracting fixpoint as long as the absolute value of all the eigenvalues of the $\partial_i f_j( \underline{m} = 0 )$ matrix is above 1. This defines a $T^* = \max_i |\lambda_i(\underline{\underline{J}})|$ temperatue. The only difference between $T_\text{c}$ and $T^*$ is the absolute value, hence $T^* \geq T_\text{c}$. If $T^* > T_\text{c}$, then starting the simulation at high temperature we will reach first $T^*$ and the simulation breaks down.
If the external magnetic field is finite, then $T_\text{c}$ is not defined in this sense, and $T^*$ is unknown before the simulation.
In that case we have to examine the eigenvalues of the
\begin{equation}
\frac{ \partial f_i }{ \partial m_j }_{\Big| \underline{m}(T)}
= \big( 1 - m_i(T) \big)^2 \frac{ J_{i j} }{T}
\end{equation}
matrix.
If at some $T^*$ the largest absolute value is 1, then the iteration will not converge anymore.
In practice, as we run the simulation, at some point, it breaks down, even if $\Delta T$ is very small.
To avoid this phenomenon, we ought to use a gradient-based minimization.

\section{Quantum mean field annealing}
\label{sec:quantum_mechanical_approach}

\subsection{Adiabatic theorem}
\label{sec:adiabatic_theorem}
Instead of the free energy and decreasing the temperature, we can use quantum mechanics with a time-dependent Hamilton operator and the adiabatic theorem to determine the system's ground state.
The adiabatic theorem asserts that if a quantum system is initially at ground
state, and the corresponding time-dependent Hamilton operator changes
sufficiently slowly, and there is a gap between the eigenvalue and the rest of the Hamiltonian's spectrum, then the system remains at the instantaneous ground state
\cite{Born1928, Kato1950}. A useful application of this theorem is to find the ground
state of a complicated Hamiltonian, i.e., the ground state of the initial Hamiltonian ($H_\text{i}$)
is easy to prepare, and the final operator ($H_\text{f}$) is the one whose ground state
we are interested in. In that case
\begin{equation}
\label{eq:hamiltonian_time_dep}
H(t) = \left( 1 - s(t) \right) H_\text{i} + s(t) H_\text{f},
\end{equation}
where $s(t)$ is a continuous, monotonic function, with $s(0)=0$ and $s(T_\text{A})=1$. The time $T_\text{A}$ is the annealing time, which must be large. The easiest choice for this is $s(t) = t/T_\text{A}$.

In a real physical device, like the D-Wave system \cite{johnson2011quantum}, this is useful, but for a simulation on a classical computer, it is impractical to simulate a large quantum mechanical system. For example, if we have an Ising model with $N$ spins, then the Hilbert space is $2^N$ dimensional, which becomes soon untractable. Therefore we will use a mean field approximation, which reduces the number of degrees of freedom, at the price of losing precision.

\subsection{Heisenberg model and the mean field annealing}
\label{sec:heisenberg_model}
The Heisenberg model is the quantum mechanical version of the Ising model. 
The corresponding Hamilton operator is
\begin{equation}
\label{eq:hamiltonian_final}
H_\text{f} = -\frac{1}{2} \sum_{i j} J_{i j} \sigma_i^z \sigma_j^z
- \sum_i h_i \sigma_i^z,
\end{equation}
where $\sigma^z$ is the Pauli z-matrix. Its eigenvectors are $| \uparrow \rangle = \left(\begin{smallmatrix}1\\0\end{smallmatrix}\right) $ and $| \downarrow \rangle = \left(\begin{smallmatrix}0\\1\end{smallmatrix}\right)$ with eigenvalues $1$ and $\minus 1$ respectively. If there is external, uniform magnetic field in the $x$ direction too, then there is an extra term
\begin{equation}
\label{eq:hamiltonian_init}
H_\text{i} = - \Delta \sum_i \sigma_i^x.
\end{equation}
This transverse term plays the same role as the entropic term in equation \ref{eq:variatonal_free_energy}. It is responsible for the mixing.
The initial ground state is 
\begin{equation}
|\Psi_0\rangle = \bigotimes_{i=1}^N
\dfrac{ | \uparrow \rangle + |\downarrow \rangle}{\sqrt{2}},
\end{equation}
which is a product state, and after the annealing the final state is the ground state of equation \ref{eq:hamiltonian_final} and from that we can determine the minimal energy spin configuration of equation \ref{eq:ising_def}.

For a given $H(t)$ Hamiltonian, the state of the system is governed by the Schr\"{o}dinger equation.
\begin{equation}
i \dfrac{ d }{ dt } | \Psi(t) \rangle = H(t) | \Psi(t) \rangle \qquad 
(\hbar = 1)
\end{equation}
Initially the system is at ground state: $| \Psi(t=0) \rangle = | \Psi_\text{g}(t=0) \rangle$, where
\begin{equation}
H(t) | \Psi_\text{g}(t) \rangle = E_\text{g}(t) | \Psi_\text{g}(t) \rangle.
\end{equation}
If the annealing time $T_\mathrm{A}$ is large enough, then $|\Psi(t)\rangle \approx |
\Psi_\text{g}(t) \rangle$. However, we do not have to solve the Schr\"{o}dinger
equation if we are only interested in the ground
state. The energy, as a functional, is enough.
\begin{equation}
E[ | \Psi\rangle; t ] = \langle \Psi | H(t) | \Psi \rangle 
\end{equation}
If this quantity is minimal for all $t$, then it defines a new dynamics, but
now we do not need $T_\mathrm{A}$ to be large. Even so we can use $s$ instead of $t$, and
change $s$ from 0 to 1. At $s=0$, we know the system's ground state, and
for $s>0$, we want to stay on the minimal energy state.

\begin{figure}
\centering
\includegraphics[width=0.5\textwidth]{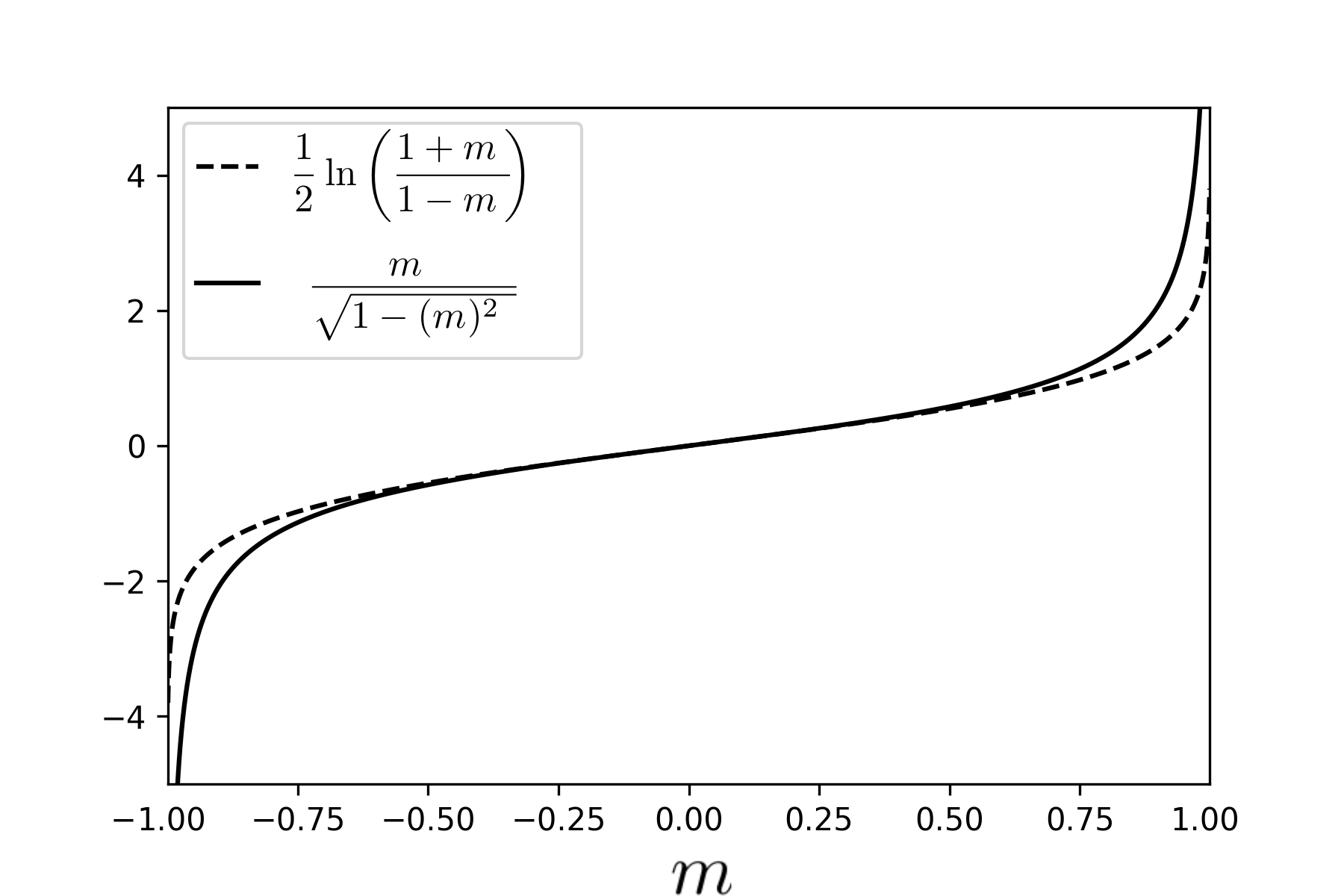}
\caption{ \textbf{ Comparison of the entropy term of statistical physical and the vertical term of quantum mechanical approach } }
\label{fig:entropy_vs_mx}
\end{figure}

So far we didn't use any approximation, which means the number of degrees of freedom is still large. To reduce it we use the mean field approximation i.e. we look for the state vector in a product form.
\begin{equation}
\label{eq:mean_field_vec}
| \Phi \rangle =
\bigotimes_{i=1}^{N} | \phi_i \rangle =
\bigotimes_{i=1}^{N} 
\left( c_{i \downarrow} | \downarrow\rangle +
c_{i \uparrow} |\uparrow\rangle \right)
\end{equation}
with the constraint: $|c_{i \downarrow}|^2 + |c_{i \uparrow}|^2 = 1$.
A useful parametrization is
\begin{equation}
\begin{aligned}
\label{eq:quatum_MF_parametrization}
m_i^z &= \langle \Phi | \sigma_i^z | \Phi \rangle =
\langle \phi_i | \sigma^z | \phi_i \rangle = 
|c_{i \uparrow}|^2 - |c_{i \downarrow}|^2\\
m_i^x &= \langle \Phi | \sigma_i^x | \Phi \rangle =
\langle \phi_i | \sigma^x | \phi_i \rangle = 
c_{i \uparrow}^* c_{i \downarrow} + c_{i \downarrow}^* c_{i \uparrow}\\
m_i^y &=\langle \Phi | \sigma_i^y | \Phi \rangle =
\langle \phi_i | \sigma^y | \phi_i \rangle = 
-i c_{i \uparrow}^* c_{i \downarrow} + i c_{i \downarrow}^* c_{i \uparrow}
\end{aligned}
\end{equation}
with real $m_i^x$, $m_i^y$ and $m_i^z$ and with the constraint: $(m_i^x)^2 + (m_i^y)^2 +(m_i^z)^2 = 1 $. It is easy to show that $c_{i \uparrow}$ and $c_{i \downarrow}$ can be choosen to be real, and in that case $m_i^y = 0$. The energy terms are
\begin{equation}
\begin{aligned}
E_\text{f} &= \langle \Phi | H_\text{f} | \Phi \rangle = -\frac{1}{2} \sum_{i j} J_{i j} m_i^z m_j^z
- \sum_i h_i m_i^z\\
E_\text{i} & = \langle \Phi | H_\text{i} | \Phi \rangle = - \Delta \sum_i m_i^x.
\end{aligned}
\end{equation}
The parameters $m_i^x$ can be expressed as $m_i^x = \pm \sqrt{1 - (m_i^z)^2}$.
Initially $m_i^x=1$ and at the end of the process it is zero and never becomes negative, so we can choose the positive solution. The energy is now
\begin{equation}
\begin{aligned}
E( \underline{m}^z; s ) =
s \left( -\frac{1}{2} \sum_{i j} J_{i j} m_i^z m_j^z
- \sum_i h_i m_i^z \right)\\
+ (1-s) (- \Delta) \sum_i \sqrt{1 - (m_i^z)^2}
\end{aligned}
\end{equation}
This equation is very similar to the statistical physical free energy in equation \ref{eq:mean_field_free_energy}. The parameter $s$ plays the role of the temperature. The large temperature is the $s=0$ and the low temperature is the $s=1$. The relevant difference is the last terms, but their purpose is the same.
The derivative divided by $s$ is
\begin{equation}
\label{eq:quantum_mean_field_derivative}
\frac{ \partial E }{ s \partial m_i^z } = 
- \sum_j J_{i j} m_j^z - h_i
+ \frac{ 1-s }{ s } \Delta \frac{ m_i^z }{ \sqrt{1-(m_i^z)^2} } = 0
\end{equation}
Now the first term is the same as in equation \ref{eq:mean_field_gradient}, and $\frac{1-s}{s} \Delta$ is equivalent to the temperature. The entropic and the transverse term are compared in figure \ref{fig:entropy_vs_mx}. Close to the origin, they are the same, and they both diverge if $m_i$ goes to $\pm 1$, so they have the same functionality.
Various scientists used similar methods \cite{goto2015bifurcation, goto2019combinatorial, leleu2017combinatorial}. They all contain the Ising term, which is initially small, then becomes dominant, and another term responsible for the mixing, and initially large then gradually vanishes.

Since the statistical and quantum mechanical approach are equivalent we will use the latter in the following. Given $m_i^z(s=0) =0$ and we want to determine $m_i^z(s=1)$. The $\underline{m}^z(s)$ value is the solution of the
$E(\underline{m}^z; s) = \min$
equation, where we assume that the $s \mapsto \underline{m}^z(s)$ trajectory is continuous.
The derivative must be zero and because the solution is a minimum the Hesse matrix must be positive definite.
\begin{equation}
\begin{aligned}
\frac{ \partial^2 E }{ \partial m_i^z \partial m_j^z } =
- s J_{i j} +
(1-s) \Delta 
\Bigg( \frac{ \delta_{i j} }{ \sqrt{1-(m_i^z)^2} } +\\
\frac{ (m_i^z)^2 \delta_{i j} }{ (1-(m_i^z)^2)^{3/2} } \Bigg)
\succ 0
\end{aligned}
\end{equation}
Without the $\underline{h}$ external magnetic field the Ising model has a $\mathbb{Z}_2$ symmetry. That means the $m_i^z = 0$ $\forall i$ is a solution to equation \ref{eq:quantum_mean_field_derivative}, and it is a minimum as long as the smallest eigenvalue of the Hesse matrix is still above zero. For the parameter $s$ it concludes
\begin{equation}
s < \frac{ \Delta }{ \lambda_{\max}( \underline{\underline{J}}) + \Delta }
\end{equation}
where $\lambda_{\max}( \underline{\underline{J}})$ is the largest eigenvalue of the matrix $\underline{\underline{J}}$.
This is the same as saying, that the critical temperature at the statistical physics case is $T_\text{c} = \lambda_{\max}( \underline{\underline{J}})$.
We can set $\Delta$ to 1 and rescale $\underline{\underline{J}}$ so that its largest eigenvalue is also 1. Now the trivial solution holds until $s$ reaches 0.5. Practically that means it is enough to start the simulation from $s=0.5$. In the simulation, once $m_i^z(s=1)$ is known, we round it to either 1 or -1. This gives us the $\underline{S}$ spin configuration.
Since the derivative of $E(\underline{m}^z; s)$ diverges as some $m_i^z$ approaches $\pm 1$ it is advantageous to use a different parametrization:
\begin{equation}
\begin{aligned}
m_i^z(\vartheta_i) &= \cos(\vartheta_i)\\
m_i^x(\vartheta_i) &= \sqrt{1 - (m_i^z)^2} = \sin(\vartheta_i)
\end{aligned}
\end{equation}
The derivative is now
%
\begin{equation}
\begin{aligned}
\frac{ \partial E(\underline{\vartheta}; s) }{ \partial \vartheta_i } = 
s \Big( \sum_j J_{i j} \sin(\vartheta_j) \cos(\vartheta_i) 
+ h_i \sin(\vartheta_i) \Big)\\
+ (1-s) (-\Delta) \cos(\vartheta)
\end{aligned}
\end{equation}
which is regular for all $\vartheta_i$.

\section{Benchmark}
\label{sec:benchmark}

\begin{figure}
\centering
\begin{subfigure}{0.45\textwidth}
\centering
\includegraphics[width=\textwidth]{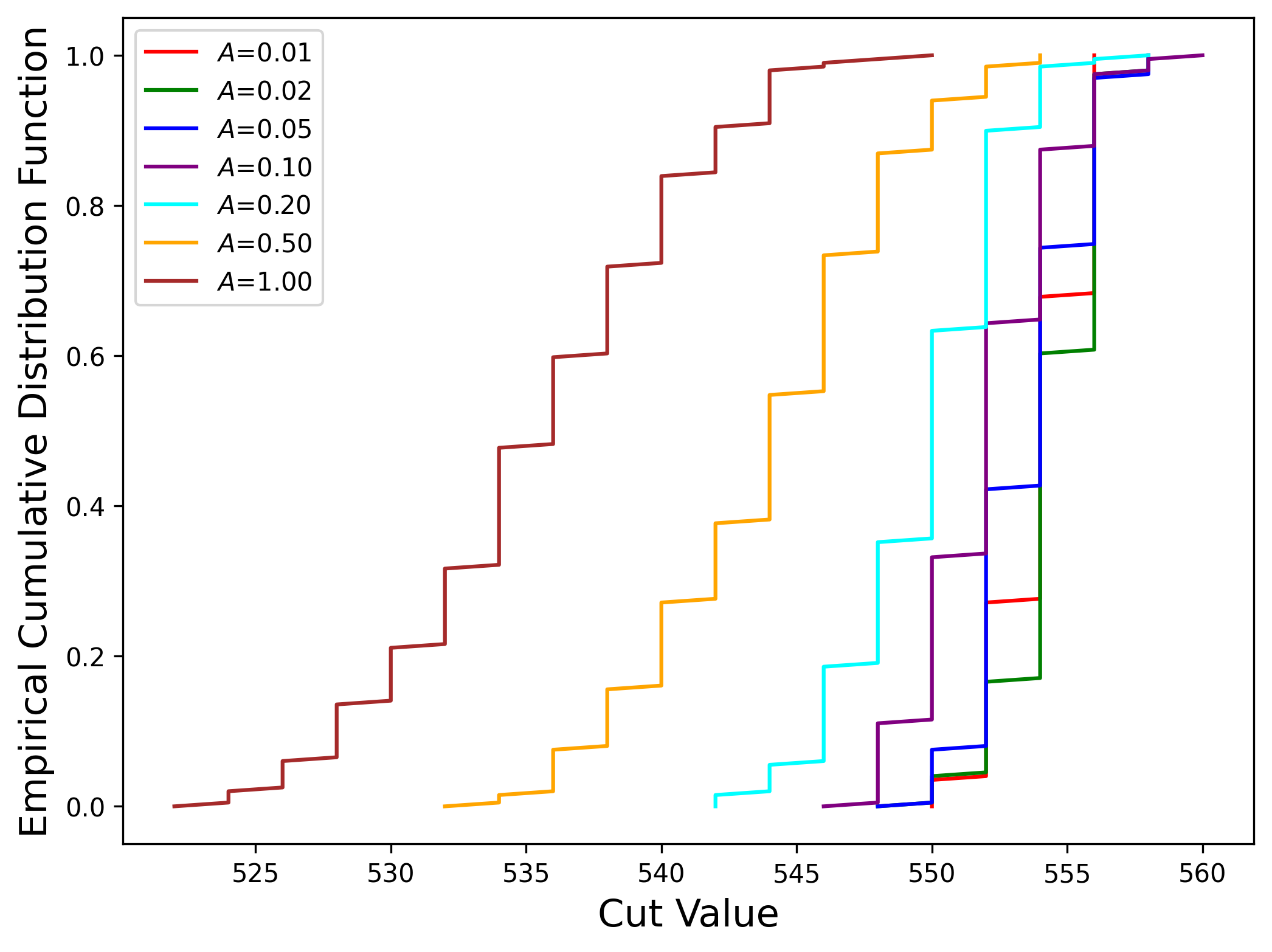}
\caption{\textbf{ Empirical Cumulative Distribution of the Cut Values of G11 for different external magnetic field amplitudes }}
\label{fig:hist_G11}
\end{subfigure}
\begin{subfigure}{0.45\textwidth}
\centering
\includegraphics[width=\textwidth]{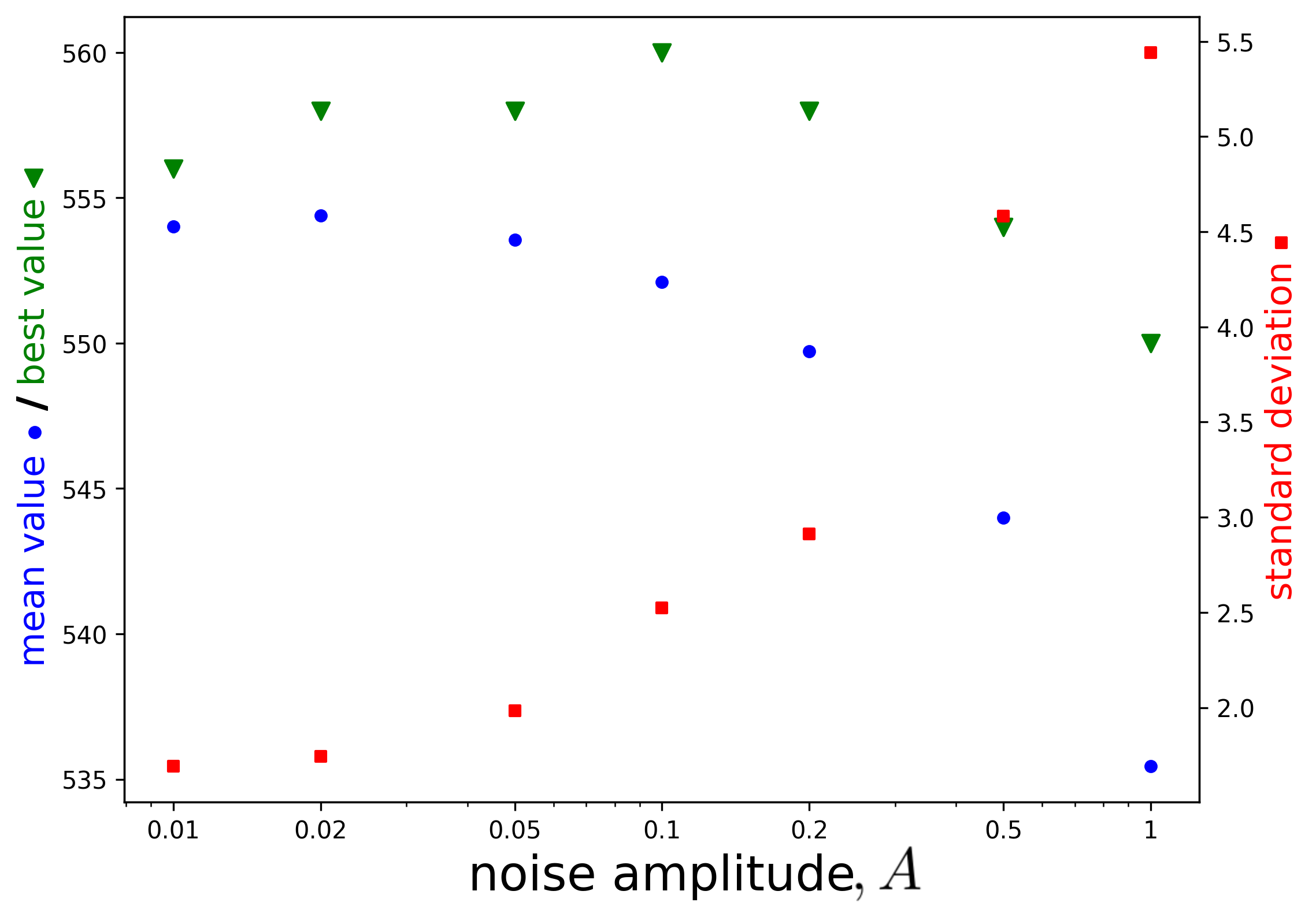}
\caption{ \textbf{ Mean value, best value and standard deviation} }
\label{fig:stats_G11}
\end{subfigure}
\end{figure}

During the simulation, the $\mathbb{Z}_2$ symmetry is disadvantageous because, without the external field, the system remains in the $m_i^z = 0$ solution forever. Choosing only one component of $\underline{h}$ to be finite breaks this symmetry. Choosing more components to be finite makes the final result ambiguous, but that can even be useful. We can use this field as noise and run the simulation many times. One such distribution is at figure \ref{fig:hist_G11}. The examined graph was the G11 from G-set \cite{helmberg2000spectral}, where the task was to find the maximal cut. This is a random graph with 800 nodes and 1600 links. The $h_i$ components are randomly generated from the $\mathrm{unif}( -A/\lambda_{\max}, A/\lambda_{\max})$ uniform distribution, where $A$ is the amplitude. Two hundred trials were generated for all amplitudes, and the step size was $\Delta s = 0.001$. The mean value, the best value, and standard deviation are shown in figure \ref{fig:stats_G11}. For small amplitudes, we have a high average CV with a small deviation. For larger amplitudes, the average decreases, but the deviation increases, resulting in a higher maximal CV. If the amplitude is too high, it becomes unlikely to obtain a high CV. 

We tested our algorithm with other Max-Cut problems from the G-set. 
We focused on the smaller ones, i.e. the largest was the G22 graph with 2000 nodes. The results are summarized in table \ref{table:G-set_benchmark}, the last column shows the best values we could find in the literature \cite{kochenberger2013solving, shao2018simple, ma2017multiple, festa2002randomized, wang2019new, matsuda_2019}.
\begin{table}


\begin{tabular}{l|l|l|l|l|l|l}
set               & G1    & G2    & G3    & G4    & G5    & G6  \\
\hline
our best result   & 11624 & 11620 & 11622 & 11646 & 11631 & 2178 \\
best known result & 11624 & 11620 & 11622 & 11646 & 11631 & 2178 \\

\end{tabular}
\\[0.5cm]
\begin{tabular}{|l|l|l|l|l|l|l}
  G7  & G8 & G9 & G10 & G11 & G12 & G22\\
\hline
  2006 & 2005 & 2050 & 1999 & 560 &  554 & 13353\\
  2006 & 2005 & 2054 & 2000 & 564 &  556 & 13359
\end{tabular}

\caption{G-set benchmark}
\label{table:G-set_benchmark}
\end{table}
In most cases, our best result is the same as the best known; in the rest, and it is still close.

\section{Conclusion and Outlook}
We have shown that the statistical physics and the quantum mechanical approach are equivalent during the mean field annealing. This is because the entropic term in the free energy and the energy of the transverse term in the Heisenberg model is very similar, as we have seen in figure \ref{fig:entropy_vs_mx}. We discussed that using the self-consistent equation during the mean field annealing is impractical because the iteration might not converge, but this problem does not exist if we use a gradient-based minimization.
Using the mean field annealing, we have solved some of the famous G-set problems with good results.
This approximation is the most straightforward version of the variational methods, and it follows that with more sophisticated approximations which take into account the correlation between the spins, we might achieve even higher Cut Values, but that also means that they will have more degrees of freedom.

\acknowledgments
This work was supported by NKFIH within the Quantum Technology
National Excellence Program (Project No. 2017-1.2.1-NKP-2017-00001) and
within the Quantum Information National Laboratory of Hungary, by the
ELTE Institutional Excellence Program (TKP2020-IKA-05) financed by the
Hungarian Ministry of Human Capacities, and Innovation Office (NKFIH)
through Grant No. K134437.

\bibliography{references.bib}
\bibliographystyle{ieeetr}
\end{document}